\documentclass[prl,aps,showpacs]{revtex4} 
\usepackage{revsymb,amsmath}
\usepackage{graphicx}
\usepackage{bbm}
\begin{document}
\author{Michael Schulz }
\affiliation{Universit\"at Ulm\\D-89069 Ulm Germany}
\email{michael.schulz@uni-ulm.de}
\author{Steffen Trimper}
\affiliation{Institut f\"ur Physik,
Martin-Luther-Universit\"at,D-06099 Halle Germany}
\email{steffen.trimper@physik.uni-halle.de}
\title{Persistence of Quantum Information}
\date{\today }

\begin{abstract}
A two-level system is considered which may perform flip-processes by coupling to 
a classical bath represented by a stochastic field. The time evolution of the 
density matrix leads to a stochastic equation with a multiplicative noise. The corresponding 
Fokker-Planck-equation (FPE) for the probability density depends on the matrix elements of 
the underlying density operator. The solution of the FPE is parametrized in terms of a new 
conserved quantity $\alpha $, which follows from the stochastic equation. The parameter 
$\alpha$ is interpreted as a measure for the persistence of quantum information. 
The FPE exhibits an single unique stationary solution which is different from Boltzmann`s law. 
The eigenvalues of the time dependent part are calculated exactly, leading to discrete relaxation 
times characterized by two quantum numbers and the ratio of Planck`s constant and the coupling 
strength to the bath. The entropy $S(\alpha )$ is analyzed as function of the quantum number $\alpha$. 
In case of $\alpha = 1$ the system is in a pure state whereas for $\alpha \neq 1$ 
a mixed state is realized. If a second two-level system is included, immersed in the common bath, 
both noninteracting two-level systems become mutually entangled. The annealed entropy is in that context non 
extensive. 

\end{abstract}


\maketitle

\noindent There is an increasing interest in the role of macroscopic environments to our 
understanding of the basics of quantum theory. The knowledge of the implications of 
the quantum theory to other theories, especially to the statistical mechanics and the 
domain of validity has captivated scientists from the beginning of quantum description. 
In such a context, the presence of an environment is commonly thought as entanglement,  
decohering and mixing properties of quantum system. Generically, an environment is assumed to be 
a noisy reservoir or a heat bath. Whereas in common interpretation of statistical mechanics 
the heat bath is unspecified, in quantum systems a heat bath can also provide an indirect 
interaction between otherwise totally decoupled subsystems and consequently a means to 
entangle them \cite{cdkl,dvclp,bfp}. In simple example for the entanglement between two qubits 
due to the interaction with a common heat bath has been explicitly shown in \cite{b}.     
Whereas in that paper the bath is described by a collection of harmonic 
oscillators, it seems to be more reasonable to specify the bath by stochastic forces represented 
by stochastic fields. From a more general point of view we expect the bath 
should be better described  in a stochastic manner and not by deterministic forces. In the 
present paper we consider a two level system (qubits) which are able to perform flip processes 
by a coupling to classical stochastic fields. Thus we bridge the gap between quantum and 
classical probability theory. This problem is related to many other questions of quantum optics 
and quantum electronics where quantum statistical aspects arising from the intrinsic quantum
character of the system while the possible time-dependence of system parameters may be interpreted 
as the influence of classical thermal fluctuations.\\

A very simple idealization of many quantum systems is a single two-level system 
which can exist in either of two states, $\left| +1\right\rangle $ and $\left|-1\right\rangle $. 
The quantum mechanics can be represented by an $U(2)$ matrix representation with the mapping $\left|
-1\right\rangle =(0,1)$ and $\left| +1\right\rangle =\left( 1,0\right) $ and
the Pauli operators $\sigma^z, \sigma^+, \sigma^-$ in their standard 
representation. Obviously, $\sigma ^{+}$ lifts the two level system from the
ground state $\left| -1\right\rangle $ to the excited state $\left|
+1\right\rangle $ while $\sigma ^{-}$ drops the system to the ground state. The
Hamiltonian of the two level system in an external time-dependent complex
scalar field $E(t)$ can be written as%
\[
\widehat{H}=\frac{1}{2}\hbar \omega \sigma^{z}-\frac{1}{2}\hbar \eta \left[
\sigma^{+}E(t)+ \sigma^{-}E^{\ast }(t)\right] 
\]%
where the energy difference between the ground state and the excited state is
given by $\hbar \omega $ and the coupling between the two level systems and
the external field is defined by $\eta $. The Hamiltonian allows the
explicit formulation of the time evolution of the density operator $\widehat{%
\varrho }$, which is formally given by the von Neumann's equation. The 
diagonal matrix elements satisfy the following equation   
\begin{eqnarray}
i\hbar \frac{\partial \varrho _{++}}{\partial t}&=&-\frac{1}{2}\hbar \eta
E(t)\varrho _{-+}+\frac{1}{2}\hbar \eta E^{\ast }(t)\varrho _{+-}  \nonumber\\
i\hbar \frac{\partial \varrho _{--}}{\partial t} &=& \frac{1}{2}\hbar \eta
E(t)\varrho _{-+}-\frac{1}{2}\hbar \eta E^{\ast }(t)\varrho _{+-}\,,  
\label{z1}
\end{eqnarray}%
whereas the non-diagonal element obeys
\begin{eqnarray}
i\hbar \frac{\partial \varrho _{+-}}{\partial t} &=& \hbar \omega \varrho _{+-}-%
\frac{1}{2}\hbar \eta E(t)\left[ \varrho _{--}-\varrho _{++}\right]\,.
\label{z4}
\end{eqnarray}%
The other element follows from $\varrho_{+-}=\varrho _{-+}^*$. 
The four components of the density matrix are
not completely independent. The sum of the diagonal elements 
$\varrho _{++}+\varrho _{--}=\rm{const}$ corresponding to the necessary
normalization condition $\text{tr}\widehat{\varrho }=\varrho _{++}+\varrho _{--}=1$. 
The external field $E(t)$ represents the influence of a classical
environment on the quantum mechanical system. We assume, that this field may 
be expressed by a complex white noise, $E(t)=E_{0}\left[ \xi _{1}(t)+i\xi
_{2}(t)\right] $, with $\xi _{1/2}(t)$ corresponding to normalized Wiener
processes $\xi _{1/2}(t)dt=dW_{1/2}(t)$ with the noise strength $E_0$. 
Thus, the density matrix $\widehat{%
\varrho }$ describing the probability distribution of the quantum process is
itself a randomly varying quantity. On the other hand, the system of
evolution Eqs.~(\ref{z1},\ref{z4}) is an ordinary set of Langevin
equations. In other words, it should be possible to map the dynamics of
these equation onto a Fokker-Planck equation. As argued above we discuss 
the dynamics in terms of the Fokker-Planck equation (FPE). To this aim, 
the Langevin-equations 
should be reduced to a set of independent
quantities representing the four components of the density matrix. The first
step is suggested by the normalization condition and leads to the parametrization  
$\varrho _{++}=\cos ^{2}\frac{\Theta }{2}\,,\quad \varrho _{--}=\sin ^{2}%
\frac{\Theta }{2}\,,\quad \varrho _{+-}=\frac{1}{2}R e^{i\phi}\,.$
Therefore the set of the three independent Eqs.~(\ref{z1}\,,\ref{z4}) is parametrized by the angles 
$0\leq \Theta \leq \pi $ and $0 \leq \Phi \leq 2\pi $ as well as the real function $R(t)$. In terms 
of these new variables Eqs.~(\ref{z1}\,,\ref{z4}) can be rewritten as three independent stochastic equations 
of the form 
\begin{eqnarray}
\sin \Theta \,\dot \Theta &=& \eta E_{0}R  
\left[ \cos \phi \xi _{2}(t)- \sin \phi \xi _{1}(t) \right]\nonumber\\
\dot R &=& \eta E_{0}\left[ \cos \phi \xi_{2}(t)- \sin \phi \xi _{1}(t)\right] \cos \Theta\nonumber\\  
R \dot \phi &=& -\omega R -\eta E_{0} \cos \Theta \left[ \sin \phi \xi _{2}(t) + \cos \phi \xi _{1}(t)\right]\,.
\label{y3}
\end{eqnarray}%
From the first and second relation follows immediately the conservation law%
\begin{equation}
\cos ^{2}\Theta (t)+R^{2}(t)=\alpha ^{2}= \rm{const.}  
\label{x0}
\end{equation}%
This conservation law is valid for each trajectory. In the following 
studies we choose $\alpha =1$ which corresponds to a pure state which will be discussed in 
detail below. We may use Eq.~(\ref{x0}) as a constraint in order to
eliminate the second relation in Eq.~(\ref{y3}). For $\alpha =1$ Eq.~(\ref{x0}) suggest to set 
$R(t)=\sin \Theta (t)$. 
It results a coupled system of two stochastic differential equations%
\[
d\Theta =\eta E_{0}\left[ \cos \phi dW_{2}(t)-\sin \phi dW_{1}(t)%
\right] 
\]%
\[
d\phi =-\omega dt-\frac{\eta E_{0}}{\tan \Theta }\left[ \sin \phi dW_{2}(t)+ \cos \phi dW_{1}(t)\right]\,, 
\]
which may be interpreted in the sense of the Ito calculus. From here, we get 
by application of the standard formalism \cite{Gardiner} the FPE  
\begin{equation}
\frac{\partial }{\partial t}P(\Theta ,\phi ,t) = \widehat{L} P(\Theta ,\phi ,t)\,,
\label{fp1a}
\end{equation}
where the operator $\widehat{L}$ is defined by 
\begin{equation}
\widehat{L} = \eta ^{2}\frac{\partial ^{2}}{\partial \Theta ^{2}} + \frac{\eta ^{2}}{4\tan ^{2}\Theta }%
\frac{\partial ^{2}}{\partial \phi ^{2}} + \frac{\omega }{2}%
\frac{\partial }{\partial \phi } \label{fp}\,.
\end{equation}%
The noise strength $E_0$ is incorporated in the coupling parameter $\eta $. 
The last equation describes the time evolution of the probability density $P(\Theta, \phi)$ 
to find a quantum statistical density operator of a stochastically driven two level system. 
The density operator is parametrized by the two remaining degrees of
freedom $\Theta $ and $\phi $ and the initial condition is given by 
$P(\Theta ,\phi ,t_{0})=\delta \left( \Theta -\Theta _{0}\right) \delta
\left( \phi -\phi _{0}\right)\,, $
where the initial values $\Theta _{0}=\Theta (t_{0})$ and $\phi _{0}=\phi
(t_{0})$ may be directly obtained from the initial quantum mechanical state. 
Without any coupling between the quantum mechanical two level system, $\eta = 0$, 
the FPE is reduced to the simple Liouville equation 
\[
\frac{\partial }{\partial t}P(\Theta ,\phi ,t)=\frac{\omega }{2}\frac{%
\partial }{\partial \phi }P(\Theta ,\phi ,t) 
\]%
with the well known solution%
\[
P(\Theta ,\phi ,t)=\delta \left( \Theta -\Theta _{0}\right) \delta \left(
\phi -\phi _{0}+\frac{\omega }{2}t\right) 
\]%
This result corresponds to the dynamics of the density matrix%
\[
\varrho (t)=\left( 
\begin{array}{cc}
\varrho _{++}(t_{0}) & \varrho _{+-}(t_{0})\exp \left\{ -i\omega t/2\right\}
\\ 
\varrho _{-+}(t_{0})\exp \left\{ i\omega t/2\right\} & \varrho _{--}(t_{0})%
\end{array}%
\right) 
\]%
with fixed diagonal elements but periodically time-dependent non-diagonal
components. In order to solve the general Fokker-Planck equation (\ref{fp}),
we introduce the new variable $\varphi =\phi +\omega t/2$. This leads to the
elimination of the drift term in Eq.~(\ref{fp})%
\begin{equation}
\frac{\partial }{\partial t}\widetilde{P}=\eta ^{2}\frac{%
\partial ^{2}}{\partial \Theta ^{2}}\widetilde{P} + \frac{%
\eta ^{2}}{4\tan ^{2}\Theta }\frac{\partial ^{2}}{\partial \varphi ^{2}}%
\widetilde{P}  \label{fp1}
\end{equation}%
with $P(\Theta ,\phi ,t)=\widetilde{P}(\Theta ,\phi +\omega t/2,t)$. The
solution of these equation can be done by an expansion in terms of eigen-
functions. To this aim we use the ansatz 
\[
\widetilde{P}(\Theta ,\varphi ,t)=Q(\Theta ,\varphi )\exp \left\{ -\lambda
t\right\} 
\]%
and obtain the eigen-value equation%
\[
\left( 4\frac{\partial ^{2}}{\partial \Theta ^{2}} + \frac{1}{\tan
^{2}\Theta }\frac{\partial ^{2}}{\partial \varphi ^{2}} + \Lambda \right) Q(\Theta ,\varphi )=0 
\]%
with $\lambda =\eta ^{2}\Lambda /4$. The parametrization of the density matrix 
requires the periodic behavior $Q(\Theta ,\varphi )=Q(\Theta
,\varphi +2\pi )$. Therefore, the eigen functions $Q(\Theta ,\varphi )$ can
be written as 
\[
Q(\Theta ,\varphi )=S(\Theta )\exp \left\{ im\varphi \right\} 
\]%
The transformations $S(\Theta )=y\left( \cos \Theta \right) \sin ^{1/2}\Theta $ and the subsequent 
substitution $x = \cos \Theta $ leads to the equation 
\begin{displaymath}
\left( 1-x^{2}\right) y''- 2x y' + 
\left( \frac{m^{2}+\Lambda -1}{4}- \frac{1+m^{2}}{4(1-x^{2})}\right) y=0\,.
\end{displaymath}%
The two independent solutions of the last equation are given in terms of the associated Legendre functions  
$P_{\nu }^{\mu }(x)$ and $Q_{\nu }^{\mu }(x)$. It results
\begin{displaymath}
y(x)= C_{1}P_{\sqrt{\Lambda +m^{2}}/2-1/2}^{\sqrt{1+m^{2}}/2}(x)+C_{2}Q_{%
\sqrt{\Lambda + m^{2}}/2-1/2}^{\sqrt{1+m^{2}}/2}(x)\,.  
\end{displaymath}%
The eigenvalues $\Lambda $ follow from the constraint, that the probability density $P(\Theta, \phi)$ 
should not offer a singular behavior. Here, this means $ (1-x^{2})^{1/4}\,y(x)$ should not singular  
for $x\rightarrow \pm 1$, which leads to  
\begin{equation}
\Lambda_{n m} = (2n-1)^2 + 2(2n-1)\sqrt{1 + m^2} +1\quad \forall n \in \mathbbm{N}.
\label{eigen}
\end{equation}
The knowledge of the eigenfunctions and corresponding eigenvalues enables us to 
construct the initial distribution and consequently, to find out the time evolution of the system. 
Since all eigenvalues $\Lambda_{nm} $ has to be positive semi-definite, 
the asymptotic distribution $\widetilde{P}_{\infty }(\Theta
,\varphi )=\widetilde{P}(\Theta ,\varphi ,t\rightarrow \infty )$ belongs to the eigenvalue 
$\Lambda = 0$. It is easy to check, see Eq.~(\ref{eigen}) that this is relalized by 
setting $n = m = 0$. There exist only one eigenfunction with $\Lambda_{00} = 0$, namely 
$P_{\infty }(\Theta ,\phi )$ = const. for all $\Theta $ and $\varphi $. On the
other hand, the final distribution function $P_{\infty }(\Theta ,\phi )$ defines a stationary state. 
Thus, $P_{\infty }(\Theta ,\phi )$ gives rise to a constant probability current, namely%
\[
J_{\Theta }=\eta ^{2}\frac{\partial P_{\infty }}{\partial \Theta }\,,\quad J_{\phi }= \frac{\eta ^{2}}{4\tan ^{2}\Theta }\frac{%
\partial P_{\infty }}{\partial \phi }+\frac{\omega }{2}%
P_{\infty }\,. 
\]%
\noindent These equations has the solution 
\[
P_{\infty }(\Theta ,\phi )=\frac{2}{\omega }J_{\phi }= \rm{const}. 
\]%
and the constraint $J_{\Theta }=0$. The probability distribution function $P(\Theta ,\phi ,t)$ 
describes the statistical properties 
of the quantum system. The variables of the function $P$ are the matrix elements of the 
underlying density operator of the quantum model. In order to interpret this on a first view 
curious result, let us go back to Eq.~(\ref{x0}). Especially we are interested in the meaning of 
the new quantum number $\alpha $ in more detail. In the general case the density matrix reads 
\begin{equation}
\widehat{\varrho } = \frac{1}{2}\left( 
\begin{array}{cc}
1 + \cos \Theta   &  R e^{i \Phi } \\ 
R e^{-i \Phi }  &  1 - \cos \Theta  
\end{array} \right)\,.
\label{dens}
\end{equation}
So it results $\det \widehat{\varrho }= (1-\alpha ^{2})/4$ and $ 
\text{tr}\,\widehat{\varrho }\,^{2} = (1+\alpha ^{2})/2$. From here we conclude 
that $\alpha$ is a real quantity with $0 \leq \alpha \leq 1$. 
Obviously, the before analyzed case $\alpha =1$ corresponds to a pure state, 
$\text{tr}\widehat{\varrho }^{2}=1$ and $\widehat{\varrho }^{2}=\widehat{\varrho }$,
respectively. Apparently, $\alpha \neq 1$ defines a mixed state. Remarkable is that the 
degree of mixing, expressed by $\alpha \neq 1$, offers a constant value independently from 
the time evolution of the external field $E(t)$. Insofar a quantum state reveals a kind of persistence 
under the influence of a classical bath. The curiosity of the result consists of that the 
probability distribution of the quantum object is again subjected to a statistics, i. e. 
the quantum probability is itself an object of a statistics. Regarding this situation 
the preparation of the quantum object will be essential, since the information, included in 
the quantum statistical operator of a closed quantum system, is determined exclusively by the 
state of the system after its preparation. Notice that $\phi$ and $ \Theta$ can be geometrically 
interpreted as spherical coordinate. In case of an arbitrary $0 \leq \alpha \leq 1$ in Eq.~(\ref{x0}), 
the angle $\Theta $ is restricted to a region $\Theta _0 \leq \Theta \leq \pi - \Theta _0$ 
with $\cos \Theta_0 = \alpha $, see Eq.(\ref{x0}). This restricts $\Theta $ to a strip around the 
equatorial cycle of a sphere. 

\noindent Let us discuss the entropy as a measure of information about the two level system. Using the 
density matrix obtained in Eq.~(\ref{dens}) and its representation 
in the form $\widehat{\varrho } = a 1 + 2 \vec b \cdot \vec \sigma  $
with appropriate chosen scalar $a$ and vector $\vec b$ as well as a representation of an arbitrary 
function $f(a 1 + 2 \vec b \cdot \sigma)$ in terms of the Pauli matrices $\vec \sigma $ \cite{ll}, 
we get the entropy, defined as $ S=-\text{tr}\widehat{\varrho }\ln \widehat{\varrho }$:
\begin{displaymath}
S=\frac{1}{2}\left[ 2\ln
2-\left( 1+\alpha \right) \ln (1+\alpha )-\left( 1-\alpha \right) \ln
(1-\alpha )\right]\,. 
\end{displaymath}
That means the complete quantum information of the two level system
is independent from the time evolution of the classical field which drives 
the system. The background for this apparently surprising result is the
understanding of the measurement processes applied to the determination of 
the density matrix $\widehat{\varrho }$. In the present case the system 
follows its evolution equation and the entropy is completely defined by 
the initial preparation by which the degree of mixing $\alpha $ is fixed. 
A further determination of $\widehat{\varrho }$ at a later time $t_{e}$ 
is not necessary. Thus, the density matrix can here interpreted in the 
sense of Bayesian statistics as a certain degree of believe suggested 
by the initial preparation. The traditional physical interpretation of a 
probability distribution function is the Gaussian frequency concept which 
is reflected in the ensemble theory. Hereby, the density matrix is measured again 
at $t_{e}$. Since a probability function bases on a frequency distribution, 
the determination of $\widehat{ \varrho} (t_e) $ requires multiple measurements 
with identical experiments. As a consequence, we get now an averaged density matrix 
$\langle \varrho (t_e) \rangle$ which includes also the average procedure 
over the external field, i.e. 
\[
\langle \widehat{\varrho }(t_{e}) \rangle = \int \widehat{\varrho }(\Theta ,\phi
)P\left( \Theta ,\phi ,t_{e}\right) d\Theta d\phi 
\]
and the actual informational content is now given by%
\begin{equation}
S_a = -\text{tr} \langle \widehat{\varrho }(t_{e} \rangle \ln \langle 
\widehat{\varrho }(t_{e})\rangle\,,
\label{entr1}
\end{equation}
which is equivalent to the thermodynamic definition of the entropy in case of an equilibrium 
state. The index a stands for annealed, so below. With other word, the classical and quantum average procedure do not commute provided the 
the mixed state of the initially realized state can be characterized on the basis of quantum mechanics.\\ 
From this point of view, we have to distinguish between the quenched average with
respect to the external stochastic field and the annealed average. The last case is realized by the average 
procedure made in Eq.~(\ref{entr1}) while the quenched case is defined as 
$S_q = - \langle \text{tr} \widehat{\varrho } \ln \widehat{\varrho }\rangle $. In that case we obtain 
\begin{displaymath}
S_q= -\frac{1}{2}%
\left[ 2\ln 2-\left( 1+\alpha \right) \ln (1+\alpha )-\left( 1-\alpha
\right) \ln (1-\alpha )\right]\,, 
\end{displaymath}
where the average is realized by means $P(\Theta ,\phi ,t)$. The difference between 
both expressions $S_a$ and $S_q$ is a consequence of the assumed non-equilibrium character of 
prepared initial state. If we start directly from a system with a maximum mixing state, 
$\alpha =0$, and therefore from the equilibrium state, we get always 
$S_q =S_a $. The discriminate between the quenched and annealed
average procedure is not necessary, if we analyze quantities which are linearly
dependent on the density matrix, for instance the averaged spin 
$\overline{S^{z}}_q= \overline{tr\widehat{\varrho }S^{z}}= 
\text{tr}\,\overline{\widehat{\varrho }}S^{z} = \overline{S^{z}}_a\,. $ 
Thus, the measurement of primarily non-thermodynamical quantities like
internal energy or averages of spin operators is not affected by the
commutation of the classical and quantum mechanical average, while all
non-linear functionals of the density matrix, like degree of mixing or
informational content depend strongly on the order the applied average
procedures.    
\noindent Now let us generalize the model to two spins in the {\em same} classical field. 
Following the procedure as before the FPE for $P(\Theta_1 ,\phi_1 ,\Theta_2 ,\phi_2 t)$ 
has the same form as in Eq.~(\ref{fp1a}). However where the evolution operator reads 
$\widehat{L} = L_1 + L_2 + L_{12}$. The additional term $L_{12}(\Theta_1 ,\phi_1, \Theta_2 ,\phi_2,t)$ 
is originated by the presence of a common bath, represented by the same stochastic field, 
for both spins. 
As a consequence we find 
\begin{displaymath}
\left \langle \widehat{\varrho }(\Theta_1 ,\phi_1 )\otimes \widehat{\varrho }(\Theta_2,\phi_2 \right \rangle 
\neq \langle \widehat{\varrho }(\Theta_1,\phi_1 ) \rangle \otimes \langle \widehat{\varrho }(\Theta_2,
\phi_2)\rangle\,.
\end{displaymath}
Whereas the total quantum entropy $S_{\rm{tot}} = S_1 + S_2 $ is additive, the total annealed entropy, 
defined as 
\begin{displaymath}
S_{\rm{tot},a} = \text{tr}\,
\overline{ \widehat{\varrho }(\Theta_1 ,\phi_1 )\otimes \widehat{\varrho }(\Theta_2,\phi_2)} 
\ln \overline{\widehat{\varrho }(\Theta_1 ,\phi_1 )\otimes \widehat{\varrho }(\Theta_2,\phi_2)}\,,
\end{displaymath}
is non extensive. A short time expansion in an interval $\Delta t $ leads to  
\begin{displaymath}
P(t) = P(t_0)\left[1 + \frac{\eta ^2 \Delta t}{\tan \Theta_1 \tan \Theta_2}\cos(\phi_1 - \phi_2 )\right]\,.    
\end{displaymath}
Although the two spins are completely without mutual interaction the system reveals a correlation between 
both particles. The quantum spins become entangled due to the common classical bath. A complete incoherent 
quantum state becomes coherent due to the coupling at a common bath. 
Concluding the paper we have considered a quantum system immersed in a classical bath represented by a 
stochastic field. In this sense the matrix elements of the density operator are subjected 
to a statistical description. They are the variables in an underlying FPE. The solution of the 
FPE is parametrized in an additional conserved quantity $\alpha$ by which mixed and pure states can be 
distinguished. When a second two-level system is immersed in the same classical bath, the initially 
decoupled quantum systems become mutually entangled. Insofar correlated baths violate the rules of the 
classical concepts of thermodynamics.

\end{document}